\documentclass[twocolumn,showpacs,preprintnumbers,amsmath,amssymb,prl,superscriptaddress]{revtex4-1}
\usepackage[dvipdfmx]{graphicx}
\usepackage{dcolumn}
\usepackage{bm}
\usepackage{color}

\begin{document}
\preprint{APS/123-QED}
\title{Frequency and wavenumber selective excitation of spin waves through \\coherent energy transfer from elastic waves}

\author{Yusuke Hashimoto}
\affiliation{Radboud University Nijmegen, Institute for Molecules and Materials, Heyendaalseweg 135, 6525 AJ Nijmegen, The Netherlands}
\affiliation{Advanced Institute for Materials Research, Tohoku University, Sendai 980-8577, Japan.}

\author{Davide Bossini}
\affiliation{Radboud University Nijmegen, Institute for Molecules and Materials, Heyendaalseweg 135, 6525 AJ Nijmegen, The Netherlands}
\affiliation{Experimentelle Physik VI, Technische Universit$\ddot{a}$t Dortmund, D-44221 Dortmund, Germany}

\author{Tom H. Johansen}
\affiliation{Department of Physics, University of Oslo, 0316 Oslo, Norway}
\affiliation{Institute for Superconducting and Electronic Materials, University of Wollongong, Northfields Avenue, Wollongong, NSW 2522, Australia}

\author{Eiji Saitoh}
\affiliation{Advanced Institute for Materials Research, Tohoku University, Sendai 980-8577, Japan}
\affiliation{Institute for Materials Research, Tohoku University, Sendai 980-8577, Japan}
\affiliation{Advanced Science Research Center, Japan Atomic Energy Agency, Tokai 319-1195, Japan}

\author{Andrei Kirilyuk}
\affiliation{Radboud University Nijmegen, Institute for Molecules and Materials, Heyendaalseweg 135, 6525 AJ Nijmegen, The Netherlands}

\author{Theo Rasing}
\affiliation{Radboud University Nijmegen, Institute for Molecules and Materials, Heyendaalseweg 135, 6525 AJ Nijmegen, The Netherlands}

\date{\today}

\begin{abstract}

Using spin-wave tomography (SWaT), we have investigated the excitation and the propagation dynamics of optically-excited magnetoelastic waves, i.e. hybridized modes of spin waves and elastic waves, in a garnet film.
By using time-resolved SWaT, we reveal the excitation dynamics of magnetoelastic waves through coherent-energy transfer between optically-excited pure-elastic waves and spin waves via magnetoelastic coupling.
This process realizes frequency and wavenumber selective excitation of spin waves at the crossing of the dispersion relations of spin waves and elastic waves.
Finally, we demonstrate that the excitation mechanism of the optically-excited pure-elastic waves, which are the source of the observed magnetoelastic waves, is dissipative in nature.

\end{abstract}



\maketitle

The development of spintronics is attracting a lot of attention due to the scaling limits of silicon based electronics.
One of the concepts for future spintronic devices relies on the transfer of data via collective oscillations of spins~\cite{Khitun:2001bv,Khitun2008,Schneider2008,Serga2010,Lenk2011,Chumak2014a,Chumak2015}, so-called spin waves or magnons.
This approach is expected to provide novel functionalities such as multi-bit parallel processing~\cite{Khitun2008}, low-energy consumption~\cite{Chumak2015}, and quantum computation~\cite{Khitun:2001bv}.
In this framework, femtosecond laser pulses have already demonstrated great potential, given their ability in the generation, manipulation and observation of the precessional motion of electron spins, even with femtosecond period and nanometer wavelength~\cite{Kirilyuk2010,Bossini2015,Walowski2016}.
Moreover, an all-optical scheme allows real-time imaging of the photo-induced spatially propagating spin waves~\cite{Satoh2012,Au2013,Yoshimine:2014hq,Iihama2016,Yoshimine:2017hm,Hashimoto2017,Kamimaki2017,Savochkin:2017ei} and the reconstruction of spin wave dispersions~\cite{Hashimoto2017}.

In magnetic media, spin waves and lattice vibrations (phonons or elastic waves) are hybridized due to the magnetoelastic coupling~\cite{Chikazumi1997}.
In particular, when an elastic wave and a spin wave have the same frequency and wavenumber, one can observe hybridization behavior, so-called magnetoelastic waves.
The concept of a magnetoelastic wave was first suggested by C. Kittel~\cite{Kittel1958} and then extensively investigated theoretically~\cite{Schlomann1960, Zapp1971, Weiler2011, Dreher2012, Ruckriegel2014, Shen2015b} and experimentally~\cite{Eshbach1962, Eshbach1963, Strauss1965, Comstock1965, Magnetization1971, Weiler2011, Weiler2012, Afanasiev2014, Janusonis2016, Kikkawa2016}.

It has recently been discovered that magnetoelastic waves can be generated by femtosecond optical excitation via the magnetoelastic coupling~\cite{Ogawa2015}.
The optical generation of magnetoelastic waves allowed the manipulation of spin textures, such as magnetic bubbles and domain walls~\cite{Ogawa2015}.
Although in this study the excitation of the magnetoelastic waves was attributed to impulsive stimulated Raman scattering (ISRS), this interpretation is controversial since the reported excitation fluence dependence exhibited threshold behavior, which has never been observed in any previous ISRS experiment~\cite{Kirilyuk2010,Bossini2017}.

\begin{figure}
\includegraphics[width=8cm]{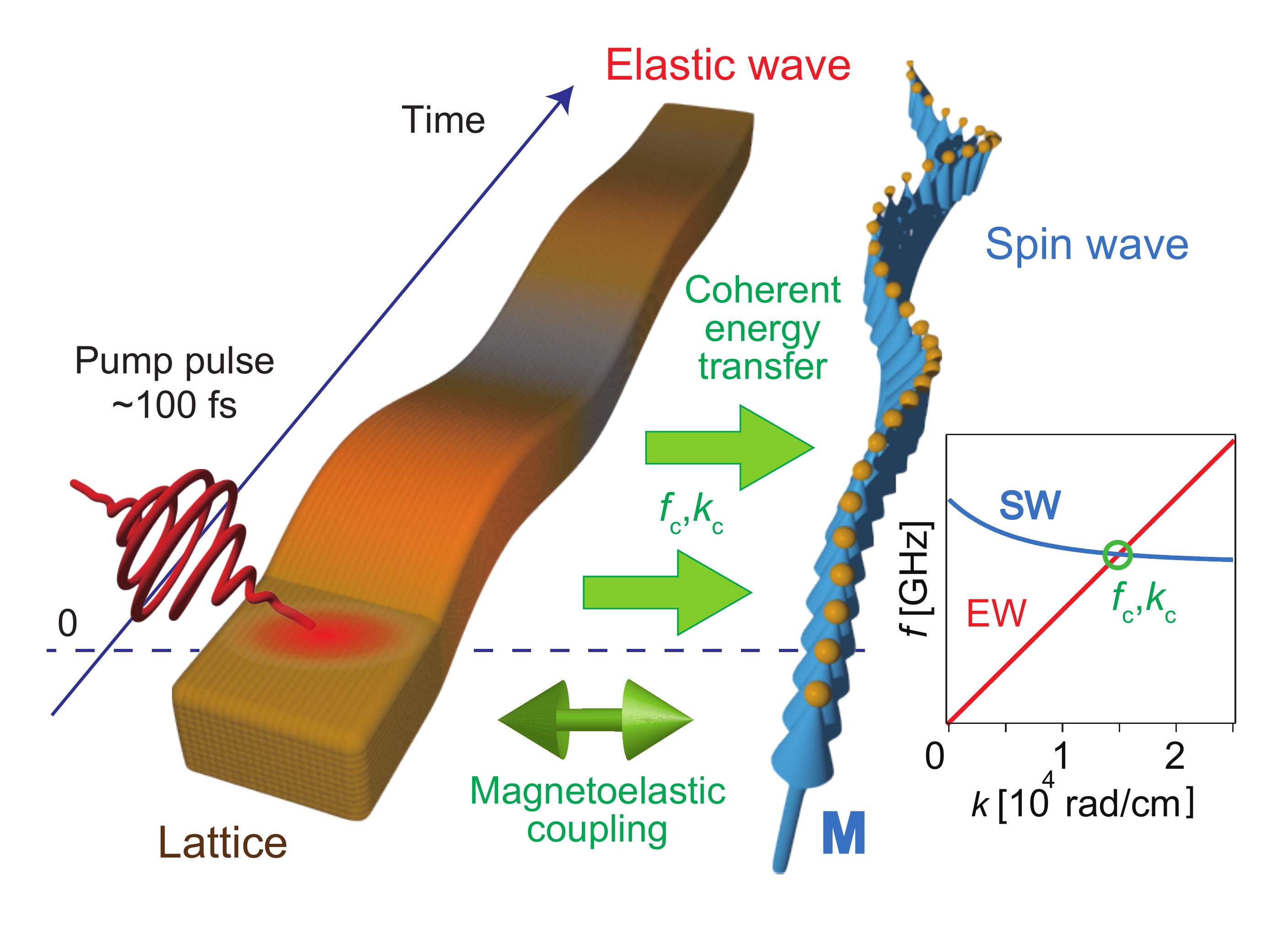}
\caption{\label{fig:concept} (color online)
Schematic illustration of the excitation of spin waves via a coherent energy transfer from optically-excited elastic waves due to the magnetoelastic coupling.
The frequency and wavenumber of the induced spin waves are determined by the crossing of the dispersion relations of spin waves (SW) and elastic waves (EW) at $f_{c}$ and $k_{c}$.
}
\end{figure}

In this study, we investigate the excitation mechanism of the optically-generated magnetoelastic waves in a ferrimagnetic garnet film by spin-wave tomography (SWaT).
By using time-resolved SWaT, we found that the magnetoelastic waves are excited by a coherent energy transfer from the optically-excited elastic waves to the hybridized magnetoelastic waves, due to the magnetoelastic coupling, as schematically shown in Fig.~\ref{fig:concept}.
Moreover, we demonstrate the manipulation of frequency and wavenumber of these magnetoelastic waves by applying an external magnetic field.

For our investigation, we chose a 4 $\mu$m thick bismuth-doped iron-garnet of Lu$_{2.3}$Bi$_{0.7}$Fe$_{4.2}$Ga$_{0.8}$O$_{12}$, which is known to give a strong magneto-optical response~\cite{Helseth2001,Helseth2002,Hansteen2004} and has a strong magnetoelastic coupling~\cite{Comstock1965}.
The sample was grown on a [001] oriented gadolinium gallium garnet substrate by liquid phase epitaxy.

\begin{figure}
\includegraphics[width=8cm]{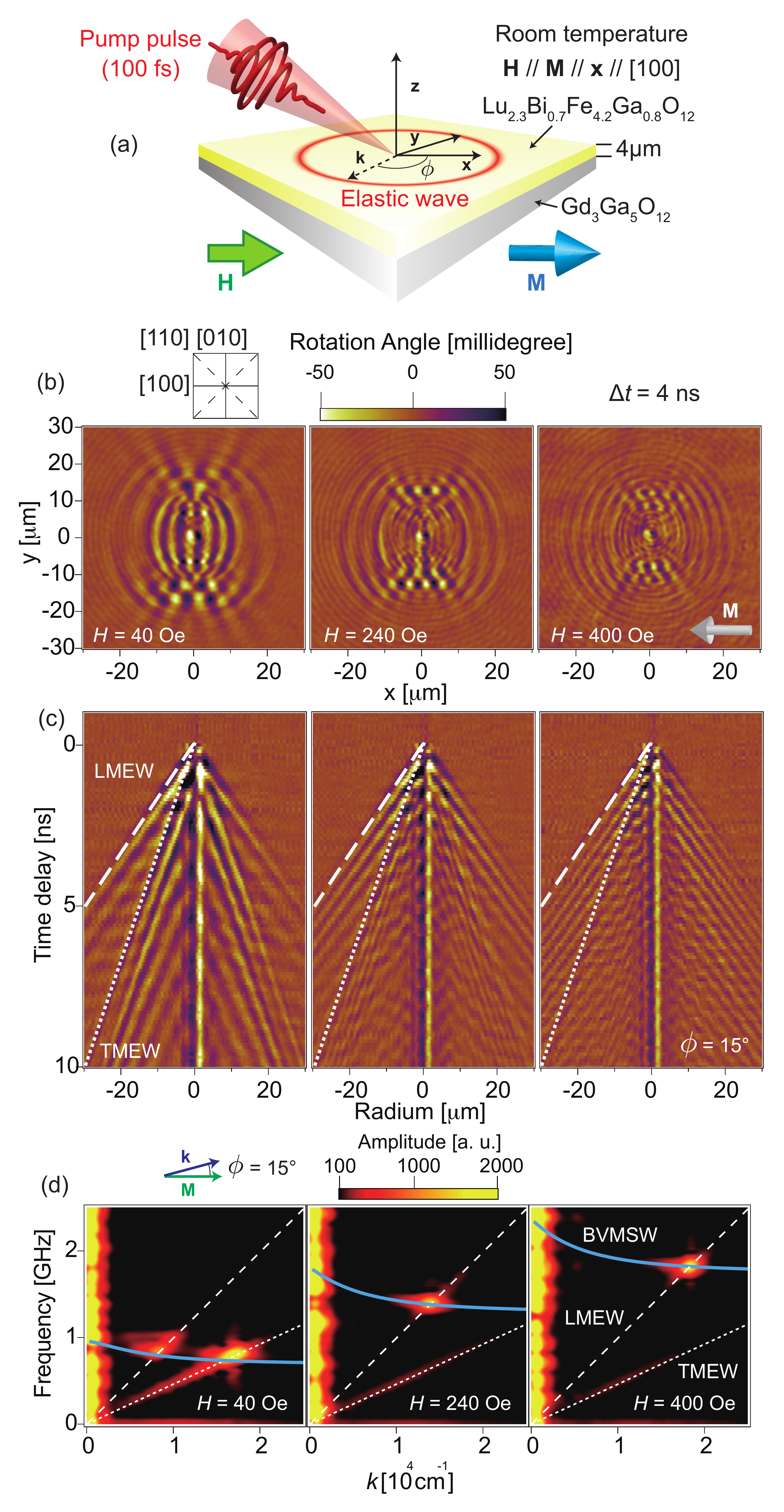}
\caption{\label{MFDep} (color online)
(a) Schematic illustration of the experimental configuration and the coordinates (x, y, z) used in this study.
The x-axis is defined along the orientation of the magnetization (${\bf M}$), parallel to the [100] axis.
An external magnetic field was applied along the x-axis.
The sample surface is in the x-y plane.
The z-axis is taken normal to the sample surface.
The wavevector of the elastic wave is defined as ${\bf k}$.
The angle between ${\bf k}$ and ${\bf M}$ is defined as $\phi$.
(b) Time-resolved magneto-optical images obtained at the time delay of 4 ns.
The intensities of the applied fields are denoted in each figure.
(c) The propagation dynamics of the optically-excited spin waves obtained by the temporal change in the images along the direction of $\phi$ = 15 degree.
The dashed and dotted lines represent the speeds of the longitudinal (LMEW) and transverse (TMEW) modes of the magnetoelastic waves, respectively.
(d) 
The magnetic field dependence of the SWaT spectra along the direction of $\phi$ = 15 degree.
The color indicates the amplitude of the spin waves with the logarithmic scale as shown in the color code.
The white dashed and dotted lines represent the dispersion relations of the LMEW and TMEW, respectively. The blue solid curves represent the dispersion relation of the backward volume magnetostatic waves (BWMSW) calculated with the Damon-Eshbach theory and the parameters obtained in Ref.~\onlinecite{Hashimoto2017}.
}
\end{figure}

The propagation of optically-excited magnetoelastic waves was observed with a time-resolved magneto-optical imaging system based on a pump-and-probe technique and a rotation analyzer method~\cite{Hashimoto2014}.
The experimental configuration is schematically shown in Fig.~\ref{MFDep}(a).
We used an 1 kHz amplified laser system generating 100 fs pulses with a central wavelength of 800 nm, which were divided into two beams: pump and probe.
Some measurements were performed tuning the wavelength of the pump beam to 400 nm, via frequency-doubling of the fundamental wavelength with a BBO crystal.
Employing an optical parametric amplifier, the wavelength of the probe beam was tuned to 630 nm, because in this spectral range the sample shows a large saturation Faraday rotation angle of (5.2 $\pm$ 0.3) degrees and high transmissivity (41 $\%$)~\cite{Helseth2001,Helseth2002,Hansteen2004}.
The pump beam was circularly polarized while all the observed waveforms shown in this study were independent of the polarization of the pump beam, including linear and circular polarizations.
The pump beam was focused to a several-$\mu$m spot in diameter on the sample surface with a fluence of 1.2 J cm$^{-2}$.
The probe beam was linearly polarized and weakly focused on the sample surface with a diameter of roughly 1 mm.
The probe fluence was 0.2 mJ cm$^{-2}$.
The transmitted probe beam was detected with a CCD camera.
The spatial resolution of the obtained images was 1 $\mu$m.
The absolute angle of the polarization plane of the probe beam was measured by using the rotation analyzer method with an accuracy of a few millidegrees~\cite{Hashimoto2014}.
The obtained images of light polarization show elastic waves through a photoelasticity~\cite{Yamazaki:2004it} and spin waves through a magneto-optical effect~\cite{Zvezdin:1997ub}, respectively.
All the waveforms discussed in this study show strong magnetic field dependences and are thus attributed to spin waves.
The obtained images of the spin waves were independent of the orientation of the polarization plane of the probe beam~\cite{Zvezdin:1997ub}, implying that the spin waves were observed through the Faraday effect.
An in-plane external magnetic field was applied along the [100] axis to control the orientation of the magnetization and to keep the sample in a single domain structure.
All the experiments discussed here were performed at room temperature.
We define in Fig.~\ref{MFDep}(a) the coordinates ($x, y, z$) and $\phi$, which is the angle between the magnetization ($\bf M$) and the wavevector of the magnetoelastic wave ({\bf k}).

Figure~\ref{MFDep}(b) shows the magneto-optical images observed under the magnetic fields of 40 Oe, 240 Oe, and 400 Oe.
The time delay between the pump and probe pulses ($\Delta t$) was set to 4 ns.
We observed spin waves propagating with complicated waveforms, showing radial and concentric structures.
Their magnetic nature was demonstrated by their strong magnetic field dependences, shown in Fig.~\ref{MFDep}(b).
Moreover, their magnetoelastic nature was indicated by their propagation speeds of 2.9 km/s (dotted line in Fig.~\ref{MFDep}(c)) and 6.2 km/s (dashed line in Fig.~\ref{MFDep}(c)), which show good agreement with the propagation speeds of the transverse and the longitudinal elastic waves~\cite{Spencer1959}, respectively.
The convincing proof of their magnetoelastic nature is obtained by the SWaT spectra, which characterizes all propagating waves by their frequency and wavenumber~\cite{Hashimoto2017}.
Figure~\ref{MFDep}(d) shows the SWaT spectra obtained from the data shown in Figs.~\ref{MFDep}(b) and~\ref{MFDep}(c).
We found that the dominant contribution to the signal appears around the crossing of the dispersion relations of elastic waves and spin waves, which is a clear indication of magnetoelastic waves.
At $\phi$ = 15 degree, pure-spin waves have negative group velocity so that the phase and the energy of the spin wave propagate in opposite directions.
This is inconsistent with the feature of the modes discussed in this study, hence ruling out the possibility to interpret our results in a purely spin-wave picture.

\begin{figure}
\includegraphics[width=8cm]{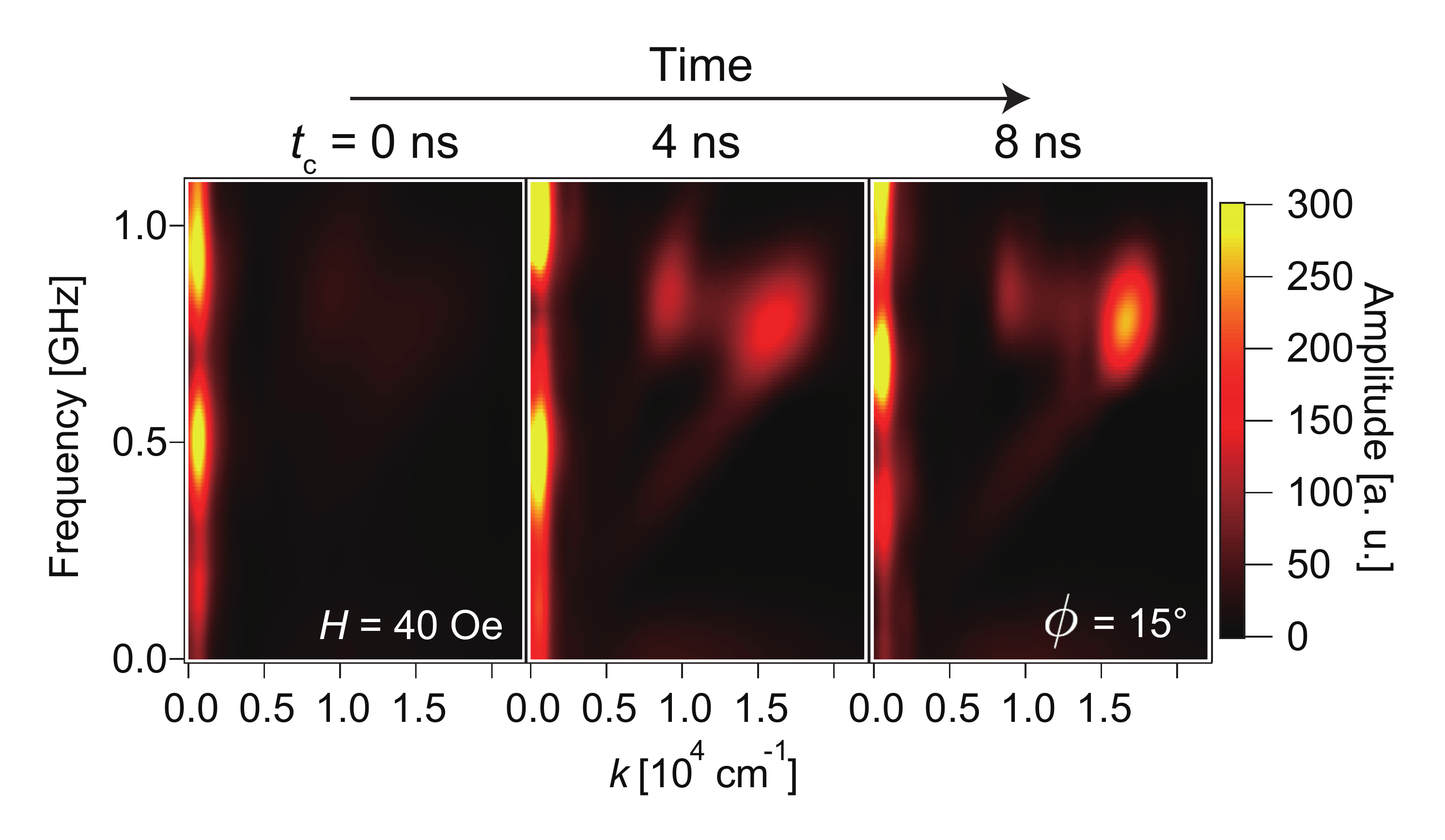}
\caption{\label{TRSWaT} (color online)
(a) Time-resolved SWaT spectra obtained by applying a Gaussian time window for the calculation of the time-Fourier transform~\cite{Hashimoto2017}.
The color indicates the amplitude of spin waves with the linear scale as shown in the color code.
We used a time window at the center time ($t_{c}$) denoted above each figures with the time width of 2.8 ns.
The same data as in Fig.~\ref{MFDep}(b) under the external field of 40 Oe was used.
}
\end{figure}

The temporal change in the amplitudes of the magnetoelastic waves is seen in time-resolved SWaT spectra obtained by applying a Gaussian time window for the calculation of the time-Fourier transform (Fig.~\ref{TRSWaT}(a))~\cite{Hashimoto2017}.
We see that the amplitudes of the magnetoelastic waves gradually increase in time.
This trend rules out impulsive stimulated Raman scattering (ISRS) as a possible excitation mechanism, since the impulsive nature of the ISRS process entails that the maximum amplitude of the magnetic oscillations occurs right after the photo-excitation~\cite{Kirilyuk2010,Bossini2017}.
Instead, we attribute the excitation of the magnetoelastic waves to the following process.
First, the optical excitation of the sample generates $pure$-elastic waves due to the absorption of the pump pulse.
Then, the $pure$-elastic waves subsequently excite magnetoelastic waves due to the magnetoelastic coupling.

In the SWaT spectra, magnetoelastic waves are observed as sharp peaks.
This feature results from the excitation mechanism of magnetoelastic waves, which is a coherent energy transfer between optically-generated elastic waves and spin waves due to the magnetoelastic coupling.
Spin waves excited by magnetoelastic coupling have a phase determined by the phase of their sources, i.e. the propagating elastic waves.
Thus, spin waves generated at different times and different positions interfere.
This interference is constructive only at the crossing of the dispersion curves of spin waves and elastic waves but destructive in other regions~\cite{Hashimoto2017}.
This process, caused by the coherence of the magnetoelastic waves, results in the selective excitation of magnetoelastic waves at a specific frequency and wavenumber.

Consequently, the identification of the excitation mechanism of the magnetoelastic waves shifts to the potential pathways of the laser-excitation of the $pure$-elastic waves.

\begin{figure}
\includegraphics[width=8cm]{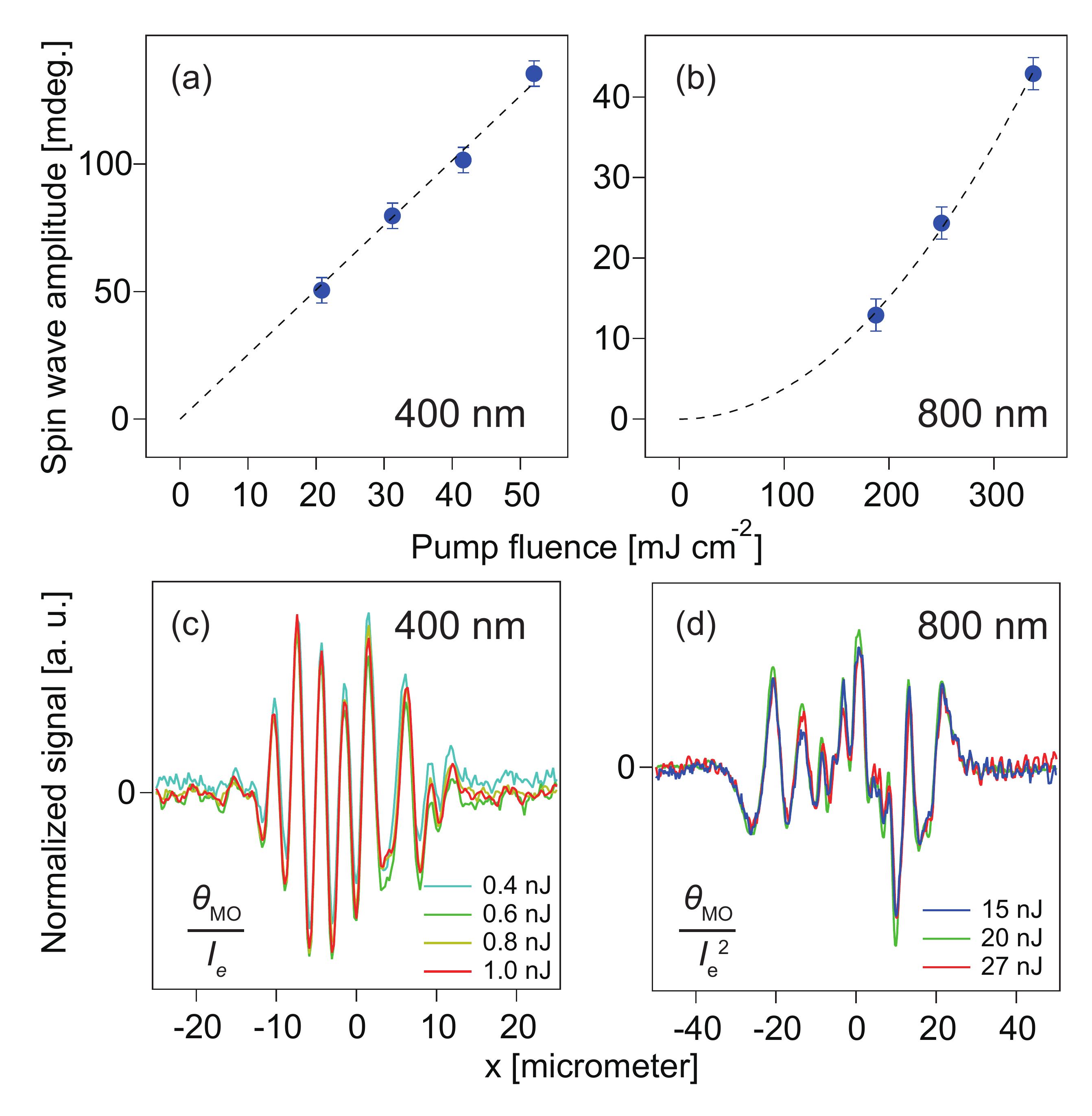}
\caption{\label{ExcDep} (color online)
The excitation fluence dependences of the amplitudes of TMEWs obtained by taking the difference of the signals of two pixels showing maximum positive and negative signals of TMEWs.
The pump wavelength was tuned to (a) 400 nm and (b) 800 nm.
The pump fluence was calculated by the focus size of the pump beam obtained by analyzing images of the photo-induced change in the transmission~\cite{Hashimoto2014}. 
The dashed lines in (a) and (b) are the fitting results of the data with linear and square functions, respectively. (c, d) The cross sections of the observed magneto-optical images ($\theta_{\rm MO}$) obtained with the pump wavelength of (c) 400 nm and (d) 800nm normalized with the linear and the square of the corresponding excitation fluence ($I_e$).
The agreement of the normalized data confirms the excitation fluence dependences indicated in (a) and (b).}
\end{figure}

The optical excitation of the pure-elastic waves is attributed to the absorption of light by the sample.
This conclusion is drawn by comparing two experiments performed with two different central pump wavelength ($\lambda_{p}$): 400 nm and 800 nm.
The LuIG film is almost transparent at 800 nm but shows large absorption around 400 nm, attributed to a charge transfer transition (CTT)~\cite{Helseth2001,Helseth2002,Hansteen2004}.
Magnetoelastic waves generated by the pump pulses with different wavelengths show similar propagation but different excitation fluence dependences of their amplitudes (Fig.~\ref{ExcDep}).
The amplitude of the magnetoelastic waves increases linearly with the excitation fluence for $\lambda_{p}$ = 400 nm (Fig.~\ref{ExcDep}(a)) but quadratically for $\lambda_{p}$ = 800 nm (Fig.~\ref{ExcDep}(b)).
These excitation fluence dependences are confirmed by comparing the cross sections of spin waves observed with different excitation fluence at the different pump wavelength.
By using the data obtained with hundreds of pixels of the CCD camera, the accuracy of the interpretation is significantly improved.
Figures~\ref{ExcDep}(c) and~\ref{ExcDep}(d) show the cross sections of the spin waves generated by pump beams with $\lambda_{p}$ = 400 nm and $\lambda_{p}$ = 800 nm, respectively.
The data were normalized with the corresponding excitation fluence dependences shown in Figs.~\ref{ExcDep}(a) and~\ref{ExcDep}(b).
The good agreements of the normalized data shown in each figure confirm the excitation fluence dependences of the spin waves shown in Figs.~\ref{ExcDep}(a) and~\ref{ExcDep}(b).
We ascribe these behaviors to the excitation fluence dependences of $pure$-elastic waves, which are generated by the absorption of the pump beam through linear absorption for $\lambda_{p}$ = 400 nm but through two-photon absorption for $\lambda_{p}$ = 800 nm~\cite{Boyd2008}.
The $pure$-elastic waves excite magnetoelastic waves due to the magnetoelastic coupling, with a strength proportional to the amplitude of the elastic waves.
It is well established that the absorption of light can trigger lattice excitations~\cite{Thomsen1986,Zeiger:1992dea}.
Disentangling the microscopic process responsible for the generation of elastic waves is beyond the scope of the present work.
The main statement demonstrated here is that the absorption of light plays a crucial role in the excitation of the magnetoelastic waves, unlike the non-dissipative ISRS picture invoked in earlier work~\cite{Ogawa2015}.
This difference may be due to the different wavelengths of the pump beams.
In the study of Ref.~\cite{Ogawa2015}, the central wavelength of the pump beam was set to 1300 nm: in this spectral range the garnet film is fully transparent so that both one- and two-photon absorption processes cannot take place, thus leaving the ISPS process as the dominant mechanism.

Elastic waves are widely used in a number of devices such as surface acoustic wave touch screens and radio frequency filters.
Thus, the mechanism demonstrated in this study, i.e. the coherent energy transfer between elastic waves and spin waves, has great potential for applications, like a frequency and wavelength selective elastic wave absorber and a spin wave transmitter.
The spectral features of these devices could be controlled by applying  a magnetic field.
This mechanism works in ferro- and ferrimagnetic materials for any elastic waves, longitudinal, transverse, and even surface acoustic waves~\cite{Dreher2012}.
Elastic waves do not have to be optically excited, but can be induced by other stimuli as well~\cite{Dreher2012}.

Our work may change the perspective concerning the role of the lattice in optically-induced spin dynamics.
So far, the majority of the experiments displayed an unavoidable and undesired scattering effect of phonons on spin waves, causing damping and relaxation of the magnetic excitation. 
Our work provides a different scenario: the lattice vibration is now one of the ideal sources of spin waves, with frequency and wavevector selectivity.

In summary, the excitation and propagation dynamics of magnetoelastic waves in a garnet film has been investigated with a time-resolved magneto-optical imaging system~\cite{Hashimoto2014}.
The dominant role played by the magnetoelastic coupling in the data has been demonstrated.
Combining our experimental investigations with the recently developed SWaT analysis, we could identify the pathways of the excitation of spins through the coherent energy transfer from the photo-induced elastic waves to the magnetoelastic waves due to the magnetoelastic coupling.
Moreover, we reveal that the excitation mechanism of the optically-generated elastic waves is dissipative in nature, being due to the absorption of light.
Finally, we discussed potential applications using the coherent energy transfer between elastic waves and spin waves, investigated in this study.

\begin{acknowledgments}
We thank Dr. T. Satoh, Dr. L. Dreher, and Prof. B. Hillebrands for fruitful discussions and Dr. S. Semin, A. van Roij and A. Toonen for their technical support.
This work was financially supported by de Nederlandse Organisatie voor Wetenschappelijk Onderzoek (NWO), de Stichting voor Fundamenteel Onderzoek der Materie (FOM), the EU-FP7 project FemtoSpin (grant no. 281043), and ERC grant agreement No 339813 (EXCHANGE). Also, this work was financially supported by JST-ERATO Grant Number JPMJER1402, and World Premier International Research Center Initiative (WPI), all from MEXT, Japan.
\end{acknowledgments}


%

\end{document}